\documentclass[twocolumn,showpacs,preprintnumbers,amsmath,amssymb]{revtex4}
\usepackage{graphicx}
\usepackage{dcolumn}
\usepackage{bm}
\begin{document}
\title{Glassy Phase in the Hamiltonian Mean Field model}
\author{Alessandro Pluchino, Vito Latora and Andrea Rapisarda}
\affiliation{Dipartimento di Fisica e Astronomia, 
Universit\`a di Catania, and INFN sezione di Catania, 
Via S. Sofia 64, 95123 Catania, Italy\\	}
\date{\today}

\begin{abstract}
We study the relaxation dynamics of a Hamiltonian system of
 $N$ fully-coupled $XY$ spins.
The thermodynamics of the system predicts a ferromagnetic and a 
paramagnetic phase.
Starting from out-of-equilibrium initial conditions, the dynamics at constant energy drives the system 
into quasi-stationary states (QSS) characterized by dynamical frustration.
We introduce the spin  polarization  as a new order parameter which allows to
interpret the dynamically generated QSS regime as a glassy
phase of the model. 
\end{abstract}
\maketitle
\vspace{0.25cm}

The Hamiltonian Mean Field (HMF) model, originally introduced
in Ref.\cite{hmf0}, has been intensively
studied in the last years for its extreme richness and
flexibility in exploring the connections between dynamics
and thermodynamics in long-range many-body systems.
In fact, on one hand the model has an exact equilibrium solution,
on the other hand, because of the presence of a kinetic energy
term in the Hamiltonian, the dynamics  can be studied 
by means of molecular dynamics  simulations \cite{hmf0,hmf1,hmf2,lh0}.
From these investigations, many new interesting 
features have emerged which are common to other systems 
with long-range interactions \cite{alfaxy,lj,nobre}.
One of the most intriguing characteristics of the 
dynamics is the existence of quasi-stationary states (QSS), 
i.e. dynamically-created states, whose lifetime 
diverges with the system size $N$ \cite{lrt_pre}. 
In such states anomalous diffusion \cite{hmf2},
non-Gaussian velocity distributions \cite{lrt_pre},
vanishing Lyapunov exponents \cite{lrt_pre},
ergodicity breaking and slow-decaying correlations \cite{mat,plr1}
have been observed.
These features have suggested a possible  application of
Tsallis generalized thermodynamics \cite{Tsa,lh1,lrt_pre,moya,cho}.

In this paper we show that the HMF model in the
QSS regime behaves similary to a glassy system. 
In fact, by means of  a new order parameter,
it is possible to characterize the dynamically generated QSS as a
thermodynamics glassy phase of the model, despite the fact that
neither disorder nor frustration are {\em a-priori} present in the
interaction.
The main idea of the paper originated from the observation
of slow relaxation  and aging \cite{mat,plr1} in the QSS regime.
Such a behavior is typical of frustrated systems, whose
prototype are spin-glasses \cite{SG}:
in these systems, the impossibility to minimize simultaneously
the interaction energies of all the couples of spins leads the
system to a very complex energetic landscape.
One might imagine it as consisting of large valleys separated
by high activation energies. Each valley contains many local
minima, i.e. metastable states, in which the system, after
quenching in his low-temperature phase, can remain trapped
for a very long time,  showing those strong memory effects
better known as aging behavior.

The HMF model describes a system of $N$ fully-coupled classical
XY spins \cite{hmf0}:
\begin{equation}
\label{spin}
\stackrel{\vector(1,0){8}}{s_i} = (cos~\theta_i,sin~\theta_i)~~~~~~i=1,...,N~~.
\end{equation}
The equations of motion derive from the following Hamiltonian:
\begin{equation}
\label{hamiltonian}
        H
= \sum_{i=1}^N  {{p_i}^2 \over 2} +
  { 1\over{2N}} \sum_{i,j=1}^N  [1-cos( \theta_i -\theta_j)]~~,
\end{equation}
where ${\theta_i}$ ($0 < \theta_i \le 2 \pi$) is the  angle 
and $p_i$ the respective conjugate variable representing the rotational 
velocity  (the mass is set equal to 1) of spin $i$.  
If we associate a particle,  moving on the unit circle, to each spin, 
the model can be seen as a system of fully-coupled {\em rotators}. 
Though the division of the potential by a factor $N$ (the so-called 
Kac's prescription) makes the Hamiltonian formally extensive \cite{lh0}, 
the latter remains nonadditive due to the long-range nature of the 
interaction \cite{lh1}. 
 
The equilibrium solution of the model in the canonical ensemble 
predicts a second-order phase transition from a high 
temperature paramagnetic (PA) phase to a low temperature 
ferromagnetic (FE) one \cite{hmf0}. 
The critical temperature is $T_c=0.5$ and corresponds to
a critical energy per particle $U_c = E_c /N =0.75$.
The order parameter of this phase transition is the modulus of
the {\it average magnetization} per spin defined as:
$M = {1\over{N}} | \sum_{i=1}^N
\stackrel{\vector(1,0){8}}{s_i} | ~~$.
Above $T_c$, in the PA phase, the spins point in different 
directions and $M \sim 0$. 
Below $T_c$, in the FE phase, all the spins
are aligned (the rotators are trapped in a single cluster) 
and $M \neq0$.

The molecular dynamics simulations at constant energy (microcanonical ensemble)
reveals interesting properties in the energy range $U=0.5-0.75$. 
In fact, starting from  of out-of-equilibrium initial 
conditions \cite{note1}, the system has an extremely   
slow relaxation to the equilibrium
and 
show the presence of meta-equilibrium  
{\it quasi-stationary states} (QSS) with the following 
properties:  

1) The temperature (calculated from the average kinetic energy)
and the magnetization assume costant values for a time $\tau_{QSS}$. 
Such values are different from the equilibrium ones and depend  
on the number of spins $N$. 

2) For large $N$, $M$ vanishes (as ${N}^{-1/6}$) and $T$ tends to an 
energy-dependent value so that  the QSS lie on the extension for $T < T_c$ 
of the high-temperature branch of the caloric curve.   

3) $\tau_{QSS}$ grows linearly with the system size $N$ \cite{hmf1}. 
For this reason the QSS regime can be interpreted as the true 
equilibrium if the thermodynamic limit is taken 
before the infinite-time limit \cite{lrt_pre}. 

4) The QSS are characterized by non-Gaussian velocity distributions  
\cite{lrt_pre}, L\'evy walks and anomalous diffusion \cite{hmf2}.  

5) The largest Lyapunov exponent vanishes and the system 
lives in a restricted  part of the {\it a-priori} accessible phase space.
Such a {\it weak-mixing} dynamics suggests a connection
with the Tsallis generalized thermodynamics \cite{lrt_pre},
but also the possibility of framing the QSS within the
so-called {\it weak-ergodicity breaking} scenario \cite{Bou}, 
typical of glassy systems. 

The last point has been recently corroborated by the discovery 
of aging in the QSS regime 
\cite{mat,plr1}. 
In the following we show how the analogy with glassy systems 
and the weak ergodicity breaking 
scenario can be made more stringent \cite{notampz} by the introduction 
of a new order parameter inspired by the microscopic dynamics of spin-glass models. 

The materials that originally were called {\it spin-glasses}
are alloys formed by a noble metal
 support (gold, silver, copper) containing randomly distributed magnetic
impurities (iron or manganese).
 Such a configuration determines a random distribution 
('quenched disorder') of the interactions: 
according to the distance between each pair of spins, the
interaction among them may be either ferromagnetic 
or anti-ferromagnetic, thus generating  frustration.
The first theoretical spin-glass model was the short-range {\it
Edwards-Anderson} (EA) model \cite{EA}. 
However, the first solvable one was the {\it Sherrington-Kirkpatrick} 
(SK) model \cite{SK1}, where the spins are coupled 
by infinite-ranged interactions independently distributed 
according to a Gaussian. 
Depending on the temperature and the parameters of the Gaussian 
distribution, the SK model shows three different phases, 
namely ferromagnetic (FE), paramagnetic (PA) and spin-glass (SG). 
Since the magnetization $M$ vanishes in the SG phase 
as well as in the PA one, an additional order parameter $q_{EA}$ 
- called {\it EA order parameter} - was  proposed \cite{EA,SK1} 
in order to discriminate between spin-glass disorder 
and paramagnetism. 
The physical meaning of this order parameter is that one of  quantifying  
the degree of freezing in the SG phase. In fact
the three phases are characterized by a different 
microscopic behavior. 
In order to get an intuitive picture of this behavior, let us 
imagine to take some snapshots of the spins configuration 
in each of the three phases \cite{Chow}. 
If a snapshot is taken at one particular time, 
one easily would be able to recognize the FE phase, 
since all the spins are aligned and frozen in their 
equilibrium position. However it would be impossible to 
distinguish between the PA and the SG phase. 
In fact in both of these phases the 
orientations of spins are random, due to the high thermal noise for
 the PA phase and to the 
quenched spatial disorder for the SG phase.  
The discrimination between these two phases is 
possible only if one takes a temporal sequence of snapshots. 
In fact in the PA phase the orientation of each spin 
at successive instants of time would be random, 
so the sequence of snapshots shows every time a different 
spatial configuration. On the other hand in the SG phase 
all the snapshots are identical, 
since each spin is frozen and retains the 
same orientation over very long periods of time. 

As previously discussed, the HMF model at equilibrium has only two phases 
(PA and FE). The main goal of this paper is to show that the 
dynamically generated QSS can be interpreted as a glassy phase 
of the model. 
For this reason, inspired by the arguments described above, 
we propose to introduce a new order parameter,  
the { \it average polarization} $p$, 
in order to measure
the extent of freezing of the system.  
The physical meaning of $p$ is related to the elementary polarizations 
$\stackrel{\vector(1,0){8}}{p_i}$, i.e. the time averages  
of the successive positions of each elementary spin vector, defined as:
\begin{equation}
\label{eq4}
\stackrel{\vector(1,0){8}}{p_i} = 
<\stackrel{\vector(1,0){8}}{s_i}(t)>= 
{1\over{\tau}} \int_{0}^\tau
{\stackrel{\vector(1,0){8}}{s_i}} (t)dt~~~~~~i=1,...,N ~~
\end{equation}
The average polarization is then obtained averaging the modulus of the elementary polarization over all the rotators:
\begin{equation}
\label{eq5}
p={1\over{N}} \sum_{i=1}^N  | \stackrel{\vector(1,0){8}}{p_i}  |~~~~.
\end{equation}
Such a new order parameter has to be compared to $M$, 
the modulus of the {\it average magnetization}, calculated as: 
\begin{equation}
\label{magnetization}
M = <M(t) > = {1\over{\tau}} \int_{0}^\tau M(t)~dt~,~~
M(t) = {1\over{N}} \left| \sum_{i=1}^N
\stackrel{\vector(1,0){8}}{s_i(t)} \right| ~.
\end{equation}
In the FE phase each elementary polarization 
vector coincides with the correspondent spin vector,
both being frozen and parallel, then the average polarization $p$
keeps a non zero value equal to $M$. 
In the PA phase the orientation of each spin vector at
every time is completely random, so this continuous
motion yields a zero value both for $M$ and $p$.
On the other hand, if the QSS correspond to a glassy-like phase of the 
model, we expect to find a zero value  for $M$, 
as in the PA phase, and a non zero value for $p$,  
as in the FE one.  
All these features are summarized in Table 1.
%
%
\begin{table}
\begin{ruledtabular}
\caption{\label{tab:table1} Values of M and p in the three phases of the HMF model}
\begin{tabular}{lcr}
   &  M  &   p   \\
\hline
\hline
Ferromagnetic phase (FE) & $\neq 0$ & $\neq 0$ \\
Paramagnetic phase  ~(PA) & 0        &      0   \\
Glassy phase             & 0        & $\neq 0$ \\
\end{tabular}
\end{ruledtabular}
\end{table}
%
%
%
%
\begin{center}
\begin{figure}
\includegraphics[width=2.8in,angle=0]{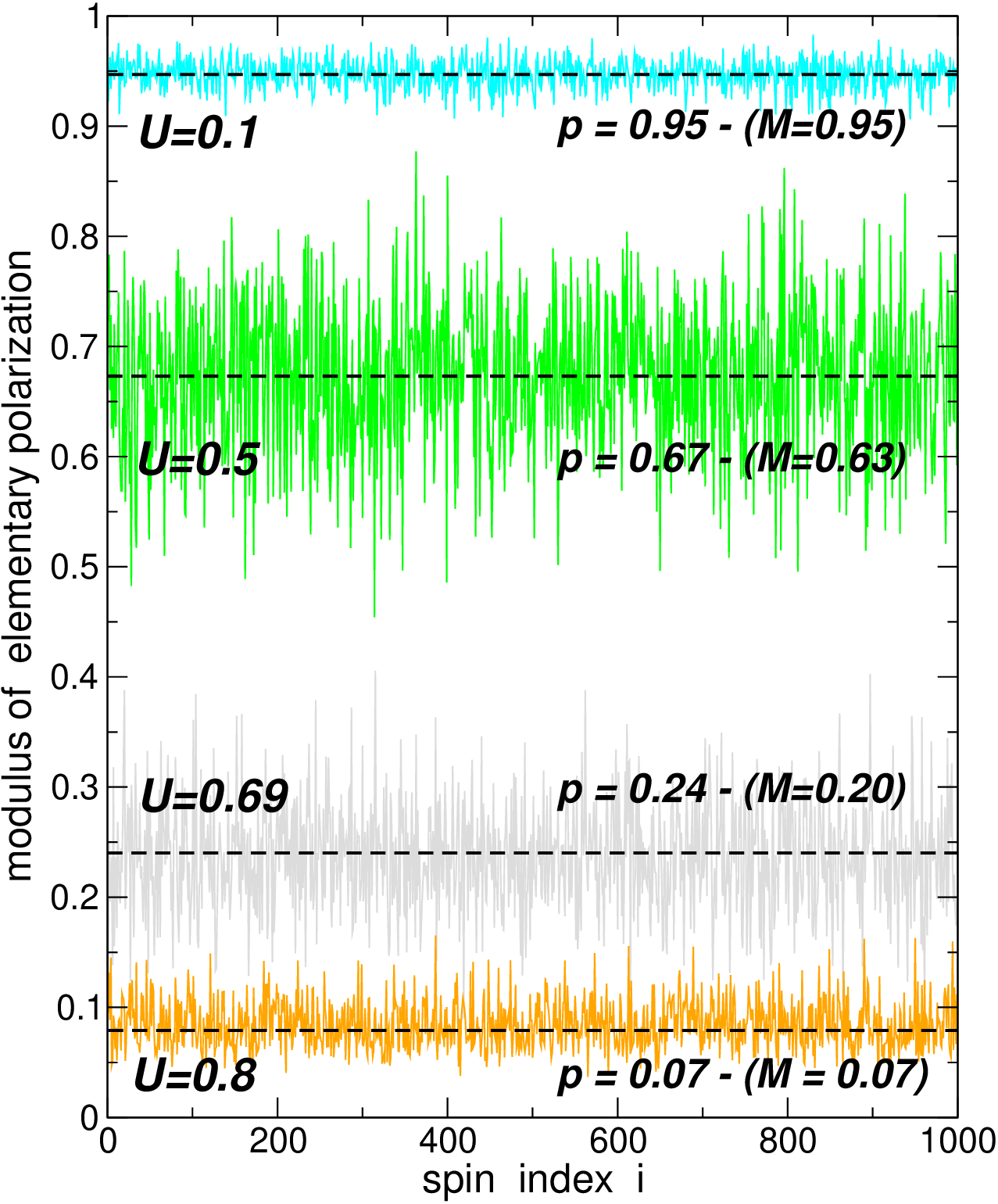}
\caption{\label{fig1}
The modulus of the elementary  polarization 
$ |\stackrel{\vector(1,0){8}}{p_i}| =  |<\stackrel{\vector(1,0){8}}{s_i}(t)>|$  
for a system with $N=1000$ and  different energies. 
The values of the average polarization $p$ 
(dashed lines) and magnetization are also reported for comparison.
Note that only for $U=0.5$ and $U=0.69$ we are in the QSS regime. In the other 
cases the system is at equilibrium}
\end{figure}
\end{center}
%
In fig.1 we show the modulus of the elementary polarization for each spin $i$. 
We consider a system of $N=1000$ spins and different energy densities. 
The values of the average polarization $p$ and the average 
magnetization $M$ are also reported in figure.   
In the simulation we have performed,  
the time averages of $p$ and $M$ 
are evaluated over an opportune time interval $\tau<\tau_{QSS}$,  
in order to stay inside the temperature 
plateau for those energy values where the QSS regime appears 
($U=0.5$ and $U=0.69$). 
In particular we have used $\tau=2000$ and a transient 
of $1000$ time units. The results do not depend significatively on $\tau$.  
As usual in molecular dynamics simulations, 
in order to make our results independent
from the specific dynamical realization,
we have also taken averages 
over a set of different realizations (events) of the same 
out-of-equilibrium initial conditions. 
As expected the two parameters $p$ and $M$ coincide and are close to $1$ at low
energy, e.g. $U=0.1$,  while both of them tend to zero for $U$ 
above the critical value $U_c=0.75$. The situation is different 
for $U=0.5$ and for $U=0.69$, two energies at which the QSS appear.  
In these cases the values of $p$ and $M$ are different:  
for $N=1000$ we have respectively 
$p=0.67,~~ M=0.63$ and $p=0.24,~~ M=0.20$.  
We have checked that the difference between 
$p$ and $M$ increases with the system size $N$.  
In particular for large $N$, in the QSS regime,  
we expect a vanishing average magnetization $M$ and an average 
polarization $p$ different from zero. 
%
%
\begin{center}
\begin{figure}
\includegraphics[width=2.5in,angle=0]{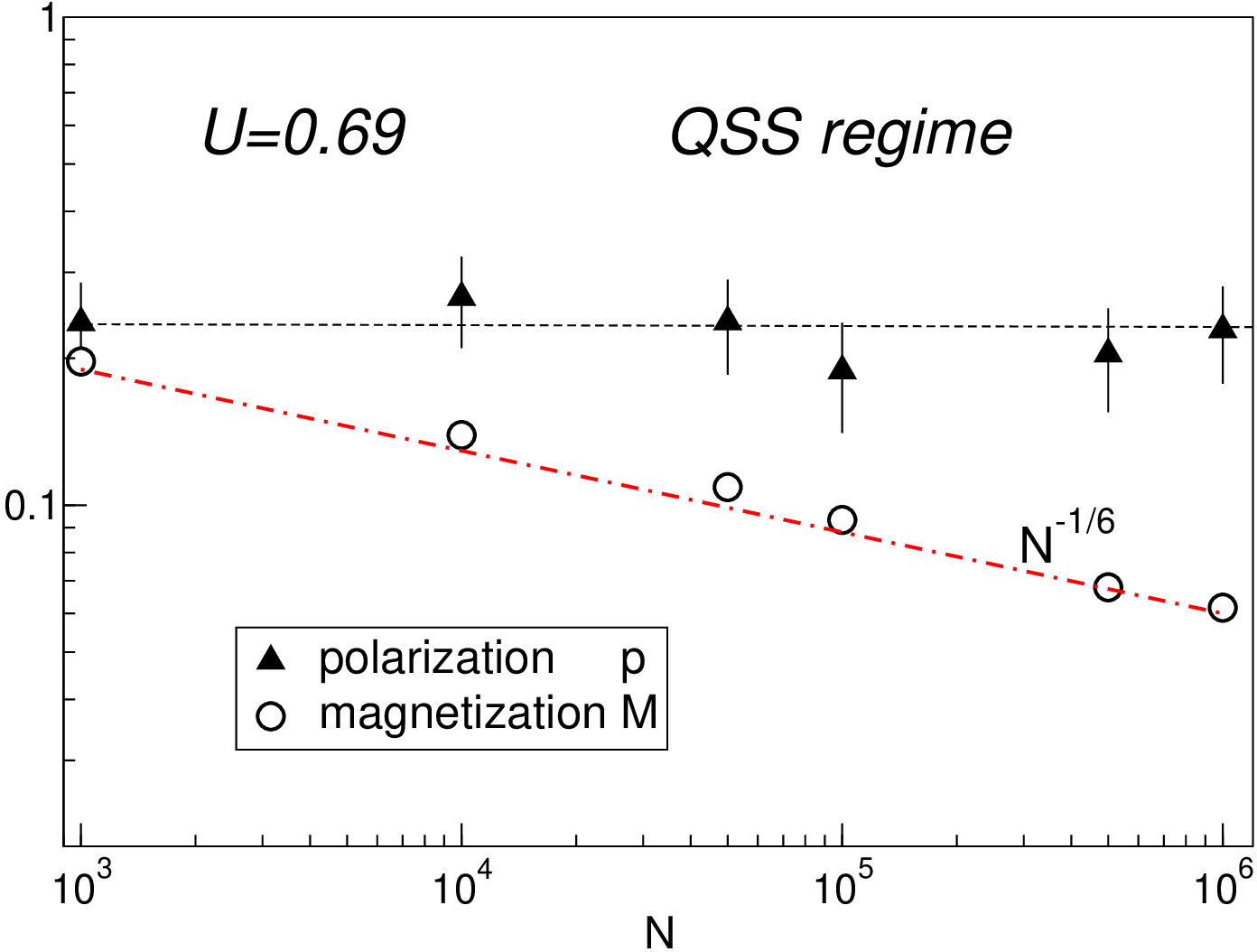}
\caption{\label{fig2} 
We plot the values of the polarization $p$ and the
magnetization $ M$ calculated in the QSS regime for $U=0.69$ 
as a function of the size $N$ of the system. 
While $p$ assumes a constant value $\sim 0.24 \pm 0.02$,  
$M$ decreases as $N^{-1/6}$. }
\end{figure}
\end{center}
%
%
%
%
\begin{center}
\begin{figure}
\includegraphics[width=2.5in,angle=0]{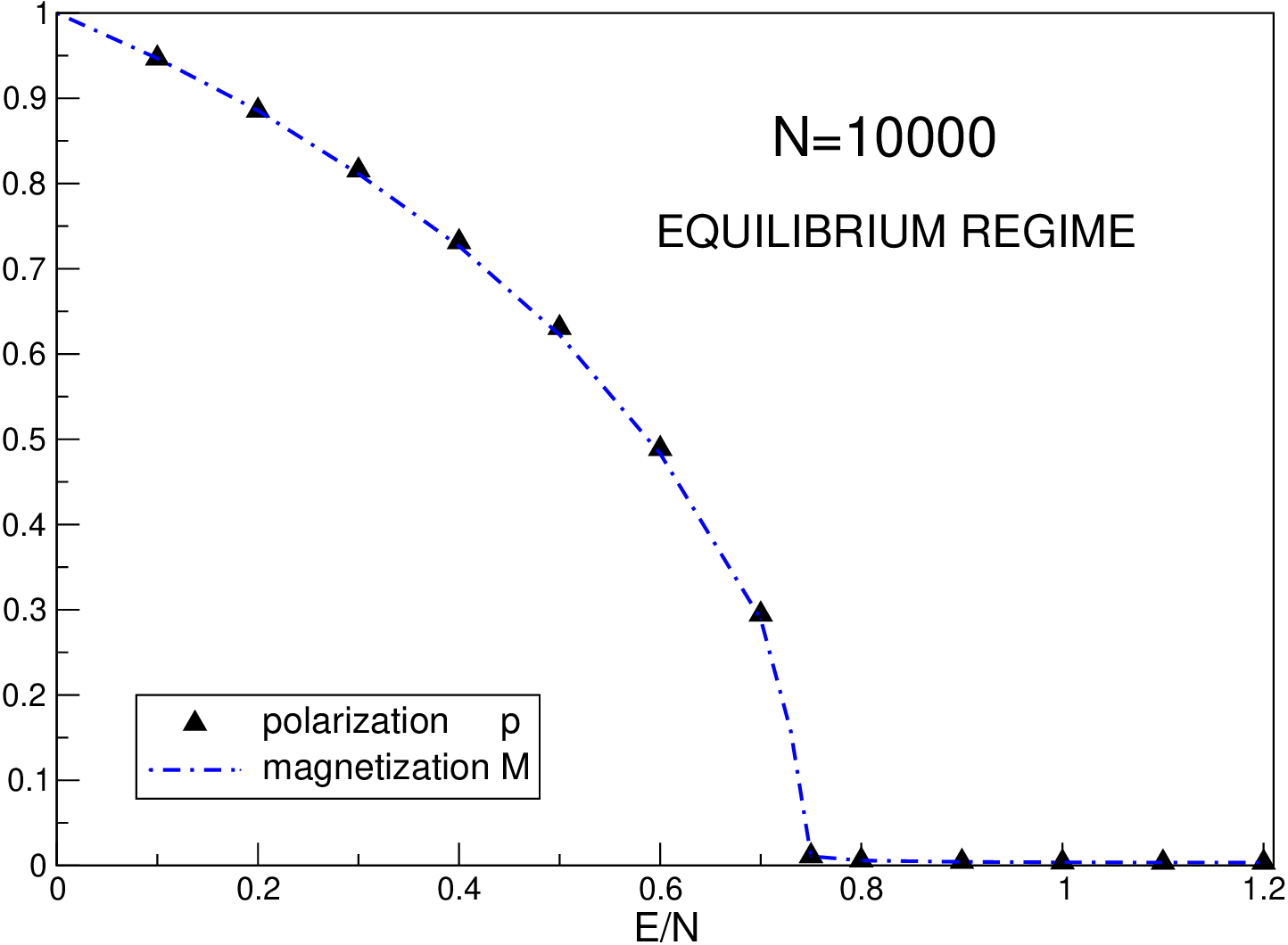}
\caption{\label{fig3} For $N=10000$, we show the polarization $p$ and magnetization $M$
vs energy per particle $U$  once the equilibrium regime has been reached.}
\end{figure}
\end{center}
%
%
In fig.2 we study the behavior of $p$ and 
$M$ with the size of the system. 
We report only the case $U=0.69$ where the anomalous effects of QSS are more evident.
As expected, while $M$ vanishes as ${N}^{-1/6}$,  $p$ 
 is independent of $N$ (within  the error) and equal to $0.24 \pm 0.02$.   

Finally in fig.3 we consider a system with $N=10000$ 
and we compare magnetization $M$ and polarization $p$  
at equilibrium for different energies.  In order 
to let the system reach equilibrium for the energy range 
$0.5\le U \le U_c$ we ran the 
simulations for a  time much larger than $\tau_{QSS}$. 
In this way every trace of metastability, and consequently also 
of the glassy phase behavior, disappears. The numerical values 
of $M$ and $p$ reported in figure coincide, in agreement with
the previous statement about equivalence between $M$
and  $p$ in the pure FE and PA phase.

Our numerical results support the interpretation of the QSS 
regime as a dynamically-created glassy phase of the HMF model. 
In the QSS regime the simulations show the formation 
of a dynamical clustering \cite{plr1}. 
The rotators feel the attraction of the 
 dynamically-generated clusters in competition within each other. 
Each rotator remains trapped  in a cluster for a while  and then eventually
succeed in escaping from it \cite{nota2}. This is also the 
cause of the anomalous diffusion and L\'evy walks observed in Ref.\cite{hmf2}.
Such a competition between the different clusters in the QSS regime
therefore realizes a  {\it dynamical frustration} that  
slows down the dynamics and prevents the system from  exploring all the 
potentially available phase space. 
Such a behavior  is also related to the aging phenomenon observed 
in refs. \cite{mat,plr1} and can be interpreted in the 
framework of the weak-ergodicity breaking scenario \cite{Bou}.  
When, at the end of the QSS regime, the system
relaxes to the equilibrium of the pure FE phase,
all the rotators concentrate in a single cluster which
rotates with the same phase of the average magnetization 
vector, i.e. $\phi={tan}^ {-1}(M_y/M_x)$  
\cite{nota3}, and all the anomalies disappear. 
%


In conclusion  the results of this paper show that the most 
remarkable features of the long-range HMF model, 
namely the dynamically-generated metastable states, can 
be interpreted as a thermodynamical glassy phase of the model. 
If the system is started sufficiently far from equilibrium,
the long-range character of the interaction produces dynamically 
 a very complex configurational landscape 
typical of glassy systems.
We have introduced  the polarization $p$
as a new order parameter 
to characterize the degree of freezing of the spins 
 due to the presence of the dynamical competition 
among  clusters in the metastable state.  
Considering that the HMF model is paradigmatic of a large class of long-range 
Hamiltonian systems, it seems  very interesting to search for  further connections
with  glassy dynamics, which  likely could  help  understanding some of the 
open problems in this field.


We thank M. Mezard,  P. Grigolini and S. Ruffo for their useful comments.

\vfill



\begin{references}
\bibitem{nota2}Although the system has a long-range interaction, one can define 
clusters by considering rotators close in angle which move with almost 
the same velocity\cite{hmf2,plr1}.
\bibitem{nota3} In order to compute
the elementary polarization of the spins,
one should consider  that
also in the QSS regime the system is always moving with a global resulting motion that can be expressed by the phase of the average magnetization $M$.
Then we have to subtract this phase from
the  spin angles. We implicitly assume this subtraction in all the simulations of the paper.
\end{references}
   \end{document}